# P-T phase diagram of iron arsenide superconductor NdFeAsO$_{0.88}$F$_{0.12}$

Alexander G. Gavriliuk[1,2,3], Viktor V. Struzhkin[1], Sergey G.Ovchinnikov[4,5], Yong Yu[1,6], Maxim M. Korshunov[4,5], Anna M. Mironovich[3], Jung- Fu Lin[6], Changqing Jin[7]

[1] *Geophysical Laboratory, Carnegie Institution of Washington, 5251 Broad Branch Road NW, Washington DC 20015, USA*

[2] *Institute of Crystallography, Russian Academy of Sciences, Leninsky pr. 59, Moscow 119333, Russia*

[3] *Institute for Nuclear Research, Russian Academy of Sciences, Troitsk, Moscow 142190, Russia*

[4] *L.V. Kirensky Institute of Physics, Siberian Branch of Russian Academy of Sciences, Krasnoyarsk, 660036, Russia*

[5] *Siberian Federal University, 79 Svobodny Prospect, Krasnoyarsk 660041, Russia*

[6] *Department of Geological Sciences, Jackson School of Geosciences, The University of Texas at Austin, Austin, Texas 78712-0254*

[7] *Beijing National Lab for Condensed Matter Physics, Institute of Physics, Chinese Academy of Sciences, P. O. Box 603, Beijing 100190, China*

# ABSTRACT

NdFeAsO$_{0.88}$F$_{0.12}$ belongs to the recently discovered family of *high*-T$_C$ iron-based superconductors. The influence of high pressure on transport properties of this material has been studied. Contrary to La-based compounds, we did not observe a maximum in T$_C$ under pressure. Under compression, T$_C$ drops rapidly as a linear function of pressure with the slope k = -2.8 ± 0.1 K / GPa. The extrapolated value of T$_C$ at zero pressure is about T$_C$ (0) = 51.7 ± 0.4 K. At pressures higher than ~18.4 GPa, the superconducting state disappears at all measured temperatures. The resistance changes slope and shows a turn-up behavior, which may be related to the Kondo effect or a weak localization of two-dimensional carriers below ~45 K that is above T$_C$ and thus competing with the superconducting phase. The behavior of the sample is completely reversible at the decompression. On the bases of our experimental data, we propose a tentative P-T phase diagram of NdFeAsO$_{0.88}$F$_{0.12}$.

The very sharp increase of $T_C$ in recently discovered family of *high*-$T_C$ superconductors (pnictides with chemical formula $RE(O_{1-x}F_x)FeAs$, see Refs.[1,2,3], where RE is the rare earth element) within a period of a few months gave hope for creating *high*-$T_C$ superconducting materials suitable for practical applications. The important feature of these new materials is very good ductility, which can make it easier to manufacture superconducting wires for electrical applications (contrary to the brittle copper oxide *high*-$T_C$ superconductors). Another very important advantage is the high value of upper critical field in iron-pnictides[4]. Change of rare earth ion or application of high pressure[2,3] indicated a substantial influence of interionic distances on $T_C$. Starting from the light rare earth element, $T_C$ increases substantially towards Nd and Sm ions[3]. Similar behavior of $T_C$ was observed due to the compression of the lattice[2]. Such similarity is not surprising because both these perturbations change interionic distances. Many efforts, both theoretical and experimental, have been put to elucidate the mechanism of superconducting (SC) pairing. The most promising candidate is the exchange of itinerant spin-fluctuations (see reviews [5-8] and references therein). Order parameter symmetry in the spin-fluctuation theories for lightly and moderately doped materials is the sign-changing s-wave (s±) [9-12], that was confirmed in many experimental studies [5]. This is quite different from the d-wave cuprate superconductors where a mechanism of pairing represents a mystery, and various models have been proposed involving antiferromagnetic fluctuations described in the framework of the t-J-type models. The conclusive explanation of the pairing mechanism in cuprates is not present yet. An interesting idea linking both classes of superconductors was outlined in the review by Grant [13], that cuprates and pnictides are dual systems which share the common features of correlated electron system with the cuprates being a strongly correlated system while the iron pnictides are more itinerant. Therefore, understanding the nature of superconductivity in iron pnictides may be a right path for uncovering the mystery of the high-$T_C$ cuprates.

The structure of $NdFeAsO_{0.88}F_{0.12}$ is P4/nmm, similar to other materials in this family [1,3,14]. Like cuprates, pnictides have a layered structure. There are Nd-O layers in which a small fraction of oxygen ions substituted by F ions to induce carriers in the conducting Fe-As layers. It has been shown[3] that oxygen vacancies in Nd-O subsystem can also serve as a source of electrons instead of adding fluorine. Fe-As conducting layers are crimpled contrary to the nearly planar Cu-O layers in cuprates. Nevertheless, some experiments have established similarities between pnictides and cuprates [15, 16]. For example, a similar 3D → 2D crossover behavior was found in the vortex structure [16]. While it is widely believed that common to all iron pnictides Fe-As layer is the playground for the formation of superconductivity, it is a big puzzle why $T_C$ varies so much among different systems (1111, 122, and 111) and even within one family. Several early systematic structure studies [17-19] suggested that the structural parameters, such as the Fe-As-Fe bond angle or the As height ($h_{As}$) above the Fe square lattice, may be important for the $T_C$ variation. First principles DFT (density functional theories) [20] and spin-fluctuation studies [21] confirmed the important role of $h_{As}$ in details of the superconducting state. Model theoretical approach is to vary $h_{As}$ and see how it affects $T_C$. In real materials, however, As height is not a free parameter and is largely controlled by the precise chemical composition. One of the best ways to study the dependence of $T_C$ on the structure is to apply a pressure sticking with one particular material. 1111 family of pnictides is promising in this respect because it contains only one Fe-As layer per unit cell and thus simpler than 122 systems. A number of high-pressure studies were carried out on this family [22-25]. Optimally doped Nd-based pnictide has a high $T_C$ among members of its family. Here we consider $NdFeAsO_{0.88}F_{0.12}$ that presents an interesting case for the high-pressure studies of the P-T phase diagram.

The high pressure-low temperature resistivity measurements were carried out with the superconductor $NdFeAsO_{0.88}F_{0.12}$ set in a diamond anvils cell. We used miniature nonmagnetic cell [26] specially designed for measurements at low temperatures and under hydrostatic conditions[27]. We used insulating gasket made of the mixture of cubic boron nitride with epoxy [28]. Sodium chloride was used as a pressure medium. The pressure was measured by the standard ruby fluorescence technique. Several ruby chips with dimensions of about 1 μm were placed into the cell along with the sample at different distances from the center of the working volume to evaluate the pressure gradient in the chamber. The pressure gradient was less than 2 GPa at all pressures. In Fig. 1 we show the image of $NdFeAsO_{0.88}F_{0.12}$ sample mounted in a diamond anvil cell [26]. The platinum foil leads with the initial thickness of about 1 micron and the width of about 10-15 microns were used for 4-probe measurements.

In Fig. 2 we show a temperature-dependent resistance measured by the four-probe method at high pressures during compression (Fig. 2a) and decompression (Fig. 2b). The resistance is normalized to $R_{min}$, the value of resistance at the point of minimum in the R(T) dependence. In Fig. 3 we show the pressure dependence of $T_C$. Its values were determined from the onset of superconducting transition in R(T) dependencies at different pressure points. The dependence is linear and can be extrapolated to about 18.4 GPa were $T_C$ drops to zero:

$$T_C = T_C(0) + k \cdot P \qquad (1)$$

where $T_C(0) = 51.7 \pm 0.4$ K, and $k = -2.8 \pm 0.1$ K / GPa.

Below $T \sim 45$ K, the resistance curve changes slope and reveals a turn-up behavior with decreasing temperature. Such a turn-up before the transition into a superconducting state has been found in several pnictides [3, 15, 25, 29]. Usually it results from a partial dielectrization due to the formation of the spin density wave (SDW) state. Nevertheless, in our case there are no evidences for the SDW state at the ambient pressure. Another reason for the turn-up may be the Kondo effect, which has been discussed recently for the $NdFeAsO_{0.7}F_{0.3}$ single crystal irradiated with $\alpha$-particles [30]. Our samples are of the same Nd-1111 family with a smaller F concentration. The other possible explanation of the turn-up may be the weak localization due to the scattering on the static randomly distributed impurities [31]. We will discuss these possibilities below.

In Fig. 4 we show the temperature dependence of the resistance measured at 25 GPa in the form $R/R_{min}$ vs. $\ln T$. There is a linear part of this dependence in the range 2.8-45 K with a minimum $R_{min}$ at $T_{min}$. We fitted the experimental data in Fig. 4 by

$$R/R_{min} = A - B \ln T \qquad (2)$$

with $A=1.15$, $B=0.043$, and $T$ being the temperature in Kelvin. Note that the behavior of $T_C$ and $R(T)$ is completely reversible at decompression.

The superconducting transition temperature $T_C$ in Nd-based sample decreases under compression following the linear pressure dependence. The extrapolated pressure corresponding to $T_C=0$ is about 18.4 GPa. Several unusual features were observed in the pressure-induced behavior of $NdFeAsO_{0.88}F_{0.12}$. The first one is the extremely rapid drop of $T_C$ under pressure (Fig. 3). The pressure slope of $T_C$ is equal to -2.8 ± 0.1 K / GPa. Similar experiments have been made by other groups on Nd-based pnictides with different oxygen deficiency [25, 32, 33]. The results of these studies are shown in the inset of Fig. 3 with our data for comparison. It is clearly seen that our samples demonstrate the maximal slope of the $T_C$ decrease under pressure. The second unusual feature is the observed anomaly in the temperature behavior of resistance before the transition into the SC state.

As to the turn-up in the resistance, one of the possible explanations can be related to the Kondo effect [3]. Magnetic impurities are inevitably present in any transition metal compounds, e.g. other rare-earth elements admixed to Nd. The $R(T)$ minimum at $T=T_{min}$ is known to result from a competition of two contributions in the total resitivity [30, 34]

$$\rho = \rho_v + c_m a \ln(\mu/T) + bT^5, \qquad (3)$$

where $\rho_v$ is the residual resistivity, the second term is the Kondo-effect contribution depending on the concentration of magnetic impurities $c_m$ and the chemical potential $\mu$, the last term stems from the electron-phonon scattering at low temperatures, $a$ and $b$ are phenomenological parameters. Minimization of Eq. (3) results in

$$T_{min} = (c_m a/5b)^{1/5}. \qquad (4)$$

The weak (power) dependence of $T_{min}$ on material parameters $c_m$, $a$, and $b$ explains the weak dependence of the $T_{min}$ on pressure in the phase diagram (Fig. 5). The parameter $a$ is related to the Kondo temperature $T_K$ as [34]

$$a \sim 1/\ln\left(\mu/T_K\right). \tag{5}$$

The success of the linear fit (2) in the experimental temperature range $T=2.8 \div 45$ K proves that $T_K \ll T$. Our estimations give $T_K \leq 10^{-1} \div 10^{-2}$ K.

The other mechanism that may result in the resistivity turn-up and $\ln T$ dependence for two-dimensional carriers is related to the quantum corrections to the conductivity in dirty metals [31]. The quasi-two-dimensional character of carriers is essential property of pnictides. Here we do not have enough information to choose whether the Kondo effect or the weak localization scenario is realized. Anyhow, both effects result in the partial localization of the carriers below $T_{min}$.

Summarizing, we have investigated the influence of high pressure on the properties of $NdFeAsO_{0.88}F_{0.12}$ high-$T_C$ superconductor. The critical temperature $T_C$ drops linearly with the slope of $-2.8 \pm 0.1$ K / GPa. The extrapolated value of $T_C$ at zero pressure is equal to $51.7 \pm 0.4$ K, close to the ambient pressure value. We suggest that at pressures higher than ~18.4 GPa the superconducting state is destabilized at all temperatures. Before the transition into the superconducting state there is an evidence of the transition into the unusual state with partial electron localization. Based on these findings, the P-T phase diagram of high-$T_C$ superconductor $NdFeAsO_{0.88}F_{0.12}$ has been proposed.

## ACKNOWLEDGMENTS

This work is supported by DOE (grant #DE-FG02-02ER45955), by the Russian Foundation for Basic Research (grants 09-02-01527-a, 09-02-00127-a, 11-02-00291-a, and 11-02-00636-a), by the Russian Ministry of Science (grant 16.518.11.7021), Presidium of RAS program "Quantum physics of condensed matter" N5.7, FCP Scientific and Research-and-Educational Personnel of Innovative Russia for 2009-2013 (GK P891, GK 2011-1.3.1-121-018 and GK 16.740.12.0731), and President of Russia (grant MK-1683.2010.2).

**Figure captions.**

**Fig. 1.** The resistivity was measured with the four-probe method on NdFeAsO$_{0.88}$F$_{0.12}$ sample in the diamond anvil cell of 300 μm culet, NaCl pressure medium, and the cBN gasket with the hole of about 80 μm diameter. The opaque NdFeAsO$_{0.88}$F$_{0.12}$ sample is located in the middle of the culet. Four Pt foil leads are connected to the opaque sample in the center of the gasket hole with NaCl medium. The pressure is about 10 GPa.

**Fig. 2.** Temperature dependence of the resistance in NdFeAsO$_{0.88}$F$_{0.12}$ sample is shown during the increase of pressure and during the decompression.

**Fig. 3.** The pressure dependence of T$_C$ in NdFeAsO$_{0.88}$F$_{0.12}$ sample can be linearly extrapolated to zero at about 18.4 GPa. Green triangle represents the ambient pressure value (from Ref. [3]). Inset shows our data (filled blue circles) in comparison with the data for FeAsNdO$_y$ from Refs. [25, 32] for y=0.6 (empty squares) and for y=0.8 (empty triangles).

**Fig. 4.** R/R$_{min}$ as a function of ln*T*.

**Fig. 5.** The P-T phase diagram of iron pnictide NdFeAsO$_{0.88}$F$_{0.12}$. The line between a usual metal above T$_{min}$ and a metallic state with the partial localization of charge carriers below T$_{min}$ was deduced from the pressure dependence of the minimum position (T$_{min}$) in the *R* vs. T curve. Solid green triangle represents the onset of the spin-density wave in pure LaOFeAs at ambient pressure (from Ref. [35]).

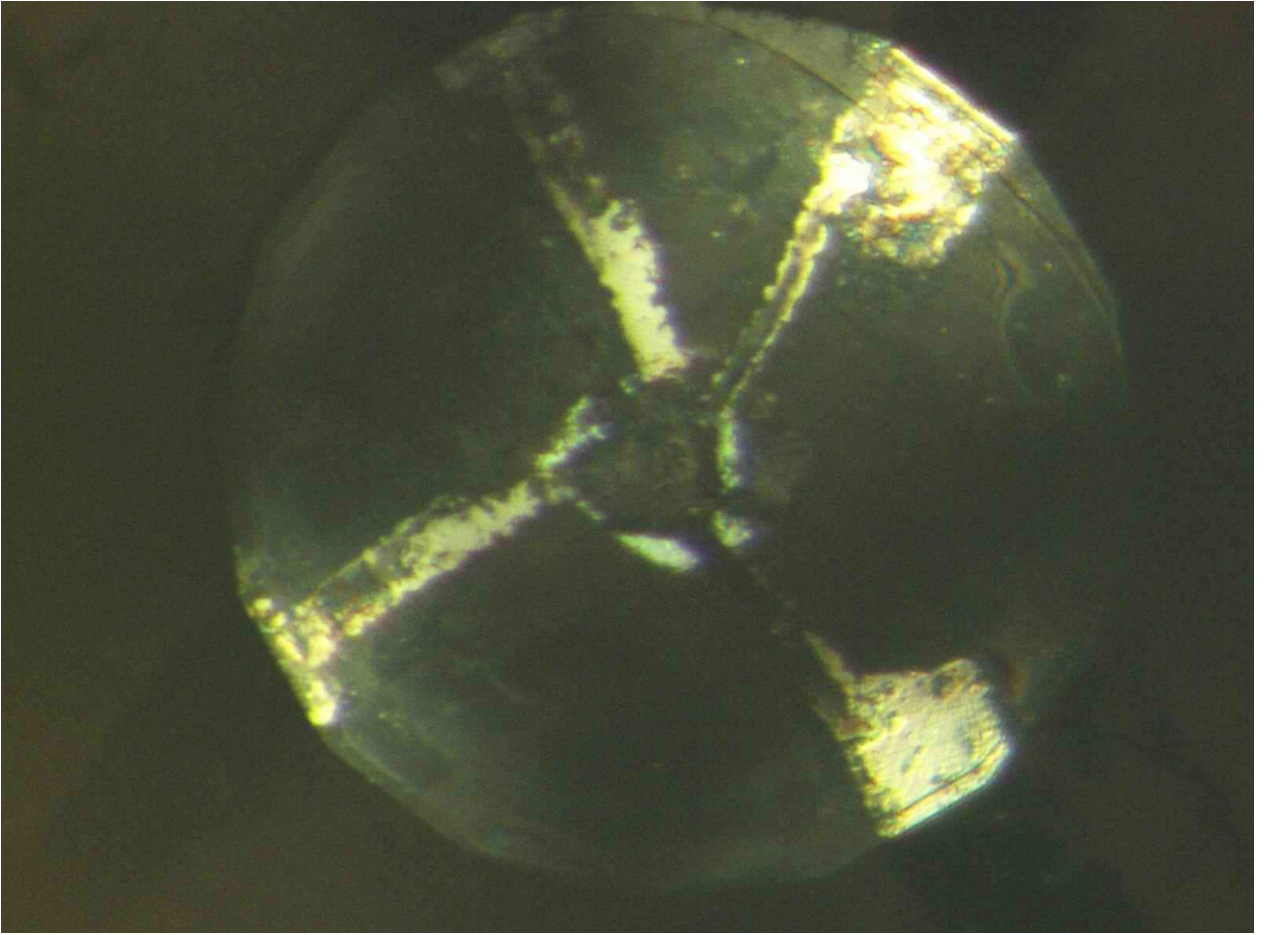

Figure 1.

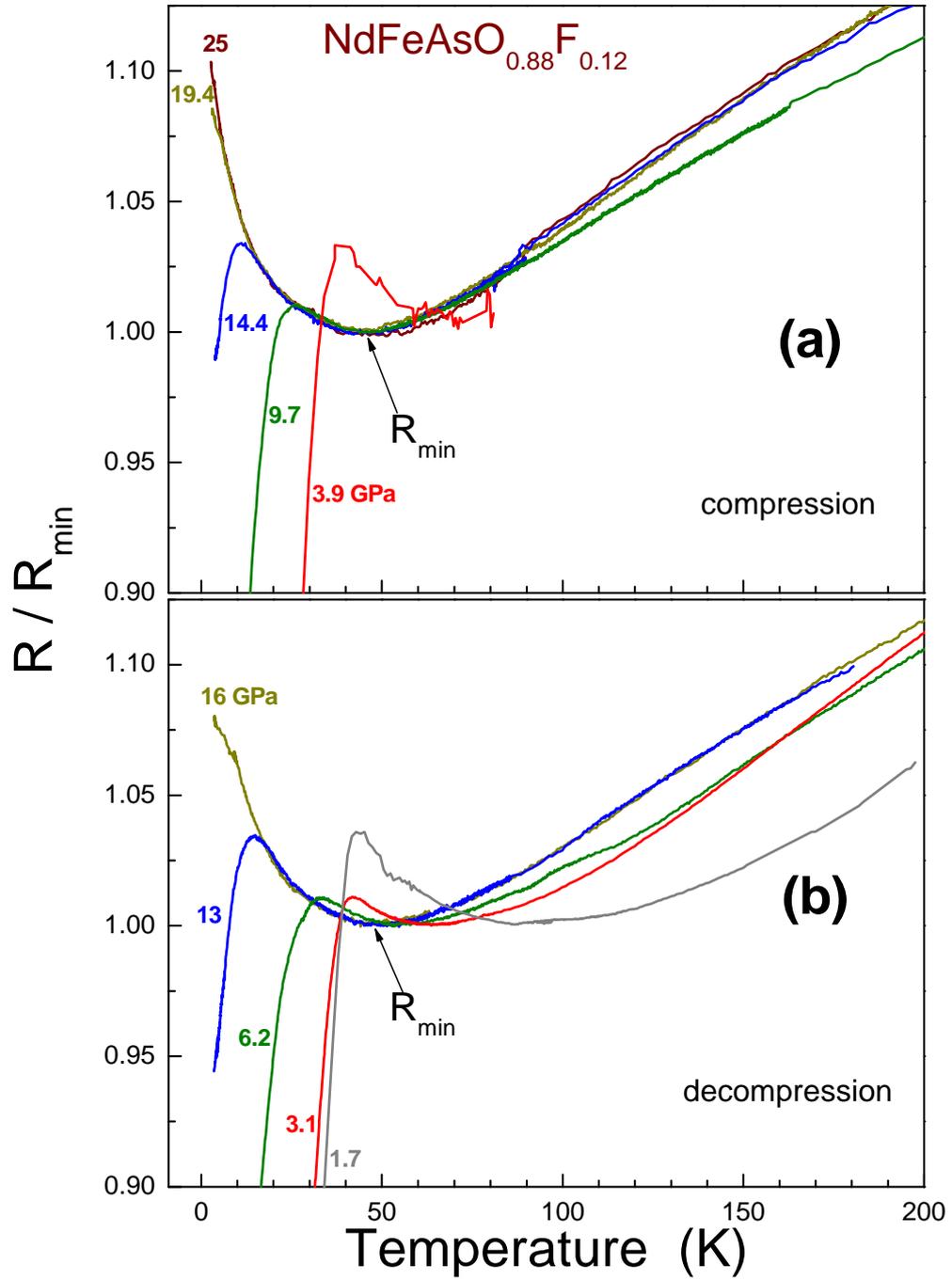

Figure 2.

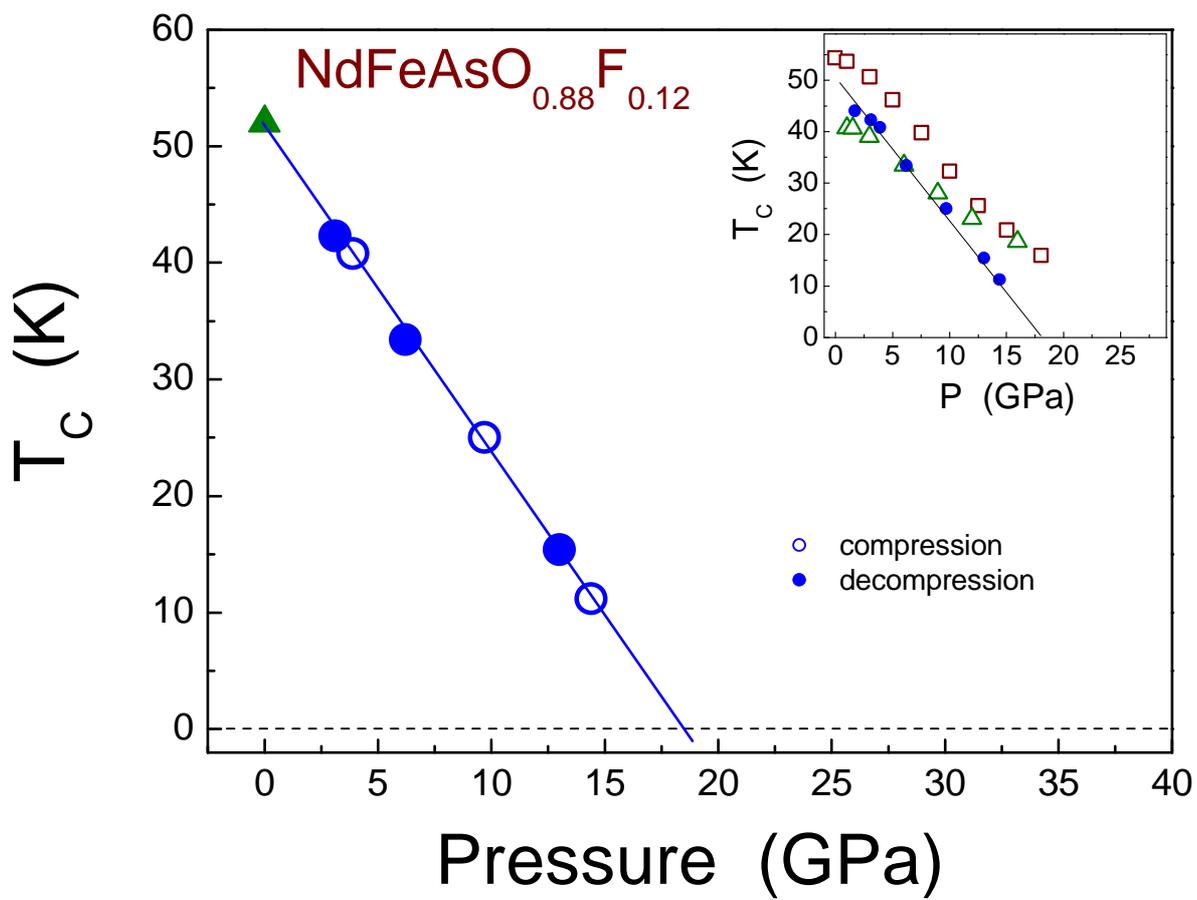

Figure 3.

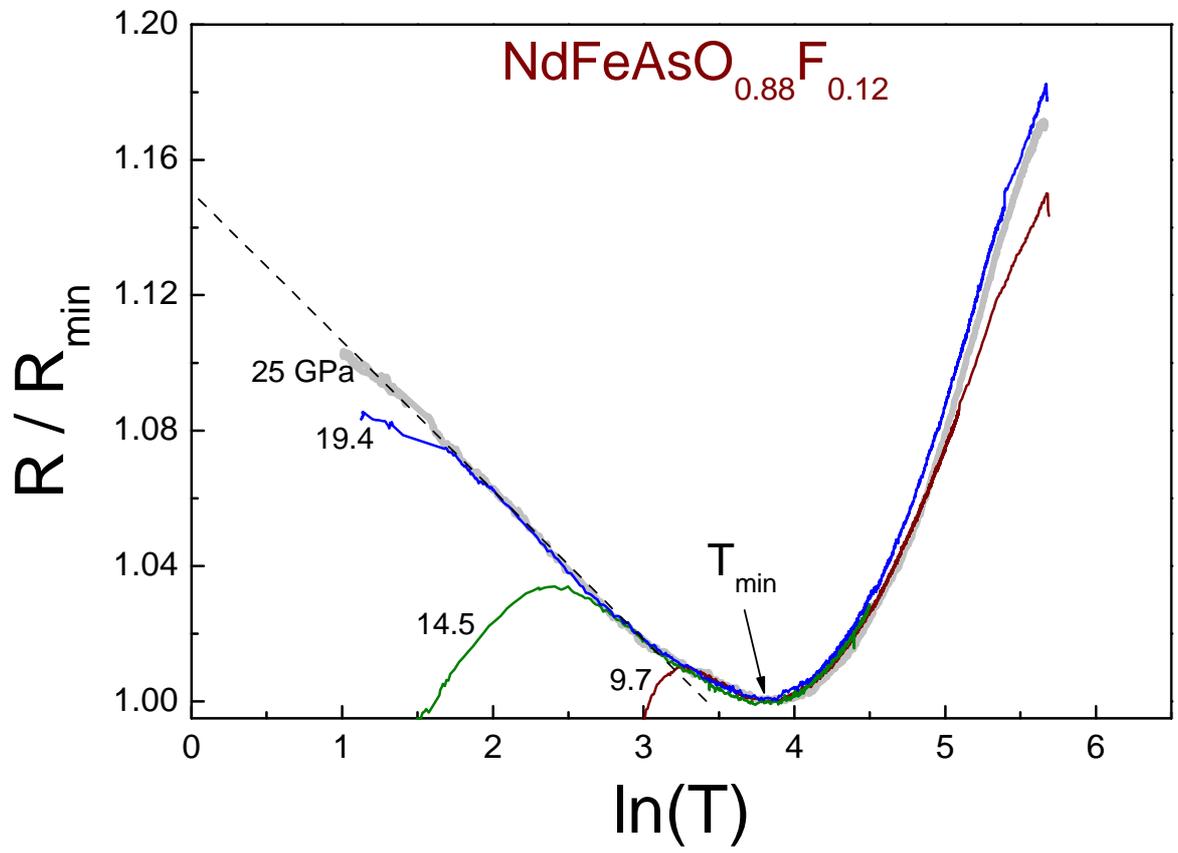

Figure 4.

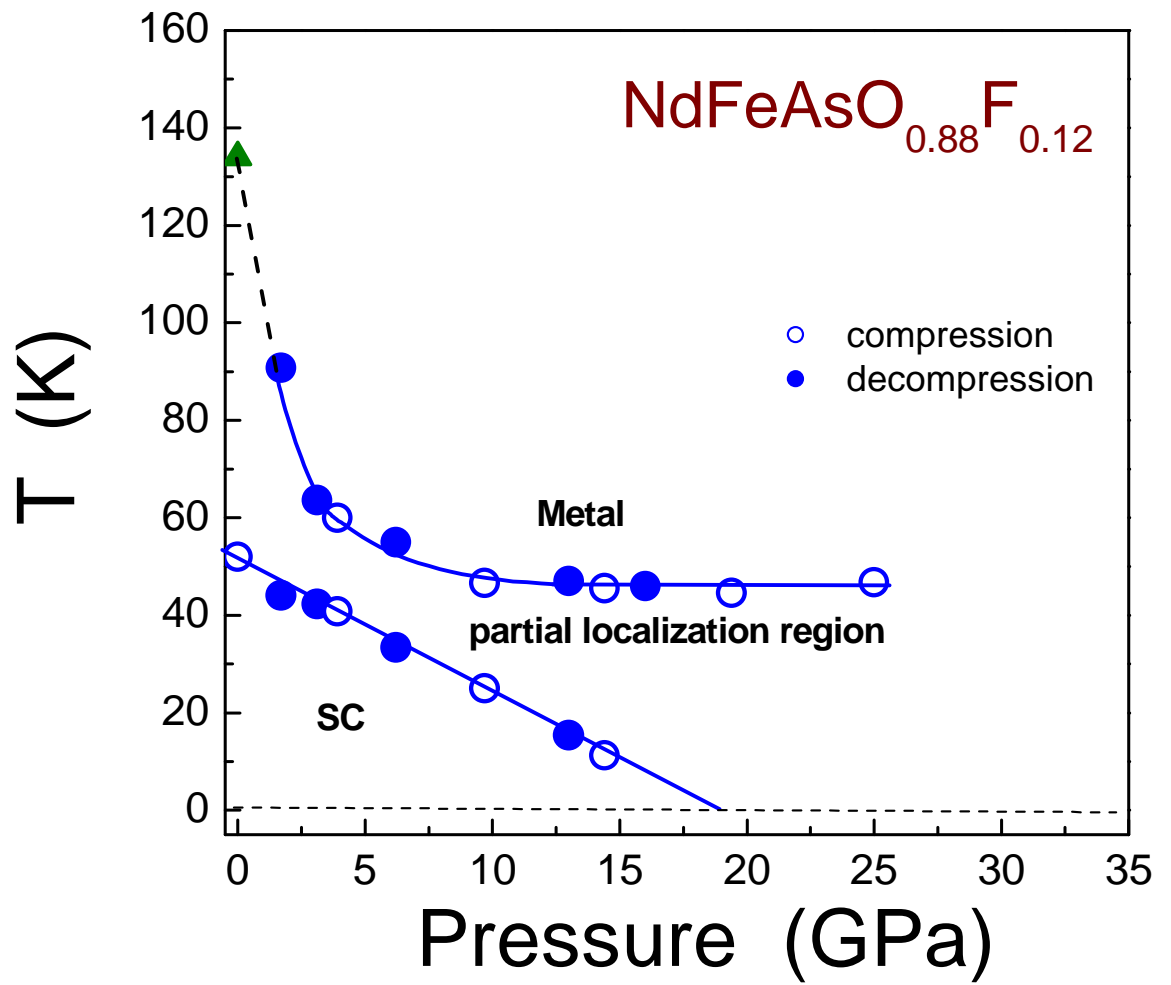

Figure 5.